\documentclass[reprint,a4paper,11pt,onecolumn,amsmath,amssymb,aps,prd,notitlepage,floatfix,nofootinbib,superscriptaddress]{revtex4-1}
\usepackage{amsmath}
\usepackage[utf8]{inputenc}
\usepackage{latexsym}
\usepackage{amsfonts}
\usepackage{graphicx,color}
\usepackage{mathrsfs}
\usepackage{epstopdf}
\usepackage{subfigure}
\usepackage[
colorlinks=true,        
citecolor=blue,         
linkcolor=blue,         
urlcolor=blue           
]{hyperref}             
\usepackage{booktabs}
\usepackage{multirow}
\usepackage{soul}
\usepackage{microtype}
\newcommand{\nc}{\newcommand}

\nc{\Eq}[1]{Eq.~\eqref{#1}}     
\nc{\Fig}[1]{Fig.~\ref{#1}}     
\nc{\Table}[1]{Table~\ref{#1}}  
\nc{\Sec}[1]{Sec.~\ref{#1}}     

\begin{document}

\title{Probing the shape of the primordial curvature power spectrum and the energy scale of reheating with pulsar timing arrays}

\author{Lele Fan}
\affiliation{School of Physics, Henan Normal University, Xinxiang 453007, Henan, China}
\affiliation{Henan Academy of Sciences, Zhengzhou 450046, Henan, China}

\author{Jie Zheng}
\affiliation{Henan Academy of Sciences, Zhengzhou 450046, Henan, China}
\author{Fengge Zhang}
\affiliation{Henan Academy of Sciences, Zhengzhou 450046, Henan, China}

\author{Zhi-Qiang You}
\email{Corresponding author: you\_zhiqiang@whu.edu.cn}
\affiliation{Henan Academy of Sciences, Zhengzhou 450046, Henan, China}

\begin{abstract}

The stochastic gravitational wave background (SGWB) provides a unique opportunity to probe the early Universe, potentially encoding information about the primordial curvature power spectrum and the energy scale of reheating. Recent observations by collaborations such as NANOGrav, PPTA, EPTA+InPTA, and CPTA have detected a stochastic common-spectrum signal, which may originate from scalar-induced gravitational waves (SIGWs) generated by primordial curvature perturbations during inflation. In this study, we explore the hypothesis that the NANOGrav signal is sourced by SIGWs and aim to constrain the shape of the primordial curvature power spectrum and the reheating energy scale using the NANOGrav 15-year data set. We model the primordial curvature power spectrum with a lognormal form and focus on the case where the equation of state during reheating is $w=1/6$, corresponding to an inflaton potential $V(\phi) \sim \phi^{14/5}$. Employing Bayesian inference, we obtain posterior distributions for the lognormal power spectrum parameters and the reheating temperature. Our results indicate a narrow peak in the primordial power spectrum ($\Delta < 0.001$ at 90\% confidence) and a lower bound on the reheating temperature ($T_{\rm rh} \geq 0.1 {\rm GeV}$), consistent with Big Bang Nucleosynthesis constraints. The best-fit SIGW energy density spectrum exhibits a distinct turning point around $f \sim 10^{-8.1}\,{\rm Hz}$, corresponding to the transition from reheating to the radiation-dominated era. This feature, combined with the sharp high-frequency decrease due to the narrow primordial power spectrum peak, offers a unique signature for probing early Universe properties.

\end{abstract}

\maketitle

\section{Introduction} \label{sec:intro}

The detection of gravitational waves (GWs) by the LIGO-Virgo-KAGRA (LVK) collaboration has been a groundbreaking achievement in the field of astronomy and astrophysics~\cite{LIGOScientific:2018mvr,*LIGOScientific:2020ibl,*LIGOScientific:2021djp}. These detections have not only confirmed the predictions of Einstein's general theory of relativity but have also opened up new avenues for studying the Universe. The LVK detectors have been instrumental in detecting GWs from binary black hole (BBH) and binary neutron star (BNS) mergers, enabling the testing of gravity in the strong-field regime~\cite{LIGOScientific:2020tif,*LIGOScientific:2021sio} and the exploration of the population-level properties of these compact objects~\cite{LIGOScientific:2018jsj,*Chen:2018rzo,*Chen:2019irf,*LIGOScientific:2020kqk,*Chen:2021nxo,*KAGRA:2021duu,*Chen:2022fda,*Liu:2022iuf,*Zheng:2022wqo,*You:2023ouk}. The successful detection of gravitational waves has paved the way for investigating the stochastic gravitational wave background (SGWB), which holds the potential to revolutionize our understanding of a wide range of cosmic mysteries, from the enigmatic nature of dark matter~\cite{Wei:2013et,*Wei:2013rea,*Du:2023zsz} to the perplexing realm of alternative theories of gravity~\cite{Chen:2014qsa,*Wu:2015naa,*Huang:2015yva,*Zhu:2015lry,*Gong:2017kim,*Chen:2024pln}, as well as other captivating phenomena at the frontiers of astrophysics and cosmology~\cite{Liu:2023qnf,*Chen:2023zkb,*Du:2024gfi,*yzq_mnras_localization,*Chen:2024jca,*Chen:2024mwg,*yzq2021gwcosmology,*He:2022ett}.

However, the sensitivity of LVK detectors is limited to high-frequency GWs, typically in the range of tens to thousands of hertz. To explore the Universe at lower frequencies, space-borne detectors such as the Laser Interferometer Space Antenna (LISA)\cite{LISA:2017pwj}, TaiJi\cite{Hu:2017mde,*Ruan:2018tsw}, and TianQin~\cite{TianQin:2015yph} have been proposed and are scheduled for launch in the near future. These detectors will be sensitive to GWs in the millihertz range, which are expected to originate from a diverse array of astrophysical sources, including massive black hole binaries and extreme mass ratio inspirals~\cite{Klein:2015hvg,*Babak:2017tow,*Gair:2017ynp,*Fan:2020zhy}. The unprecedented sensitivity of these space-based observatories will open a new window into the GW Universe, complementing the observations made by ground-based detectors and providing invaluable insights into the dynamics of massive compact objects and the evolution of galaxies across cosmic time.

Pulsar timing arrays (PTAs) are the most promising tools for probing even lower frequencies in the nanohertz range. PTAs are networks of precisely timed millisecond pulsars that can detect SGWBs~\cite{1978SvA....22...36S,*Detweiler:1979wn,*1990ApJ...361..300F}. The presence of a SGWB induces correlated timing residuals in the pulses of millisecond pulsars, which can be detected by analyzing the timing data from multiple pulsars distributed across the sky. The unique frequency range and the ability to observe the entire sky make PTAs complementary to ground-based and space-borne GW detectors, providing a comprehensive view of the GW Universe~\cite{Tiburzi:2018txc,*Kelley:2017lek,*Li:2019vlb,*Chen:2019xse,*Wu:2023pbt,*Wu:2023dnp,*PPTA:2022eul,*Chen:2021wdo,*Chen:2022azo,*Chen:2021ncc,*Wu:2021kmd,*IPTA:2023ero}.
Currently, several PTAs are in operation, including the North American Nanohertz Observatory for Gravitational Waves (NANOGrav)\cite{McLaughlin:2013ira}, the European Pulsar Timing Array (EPTA)\cite{Kramer:2013kea}, the Parkes Pulsar Timing Array (PPTA)\cite{Manchester:2012za}, and the Chinese Pulsar Timing Array (CPTA)\cite{2016ASPC..502...19L}. These collaborations have been collecting pulsar timing data for over a decade, aiming to detect SGWBs and study their properties. Recent analyses of the NANOGrav 15-year data set~\cite{NANOGrav:2023hde,*NANOGrav:2023gor}, PPTA DR3~\cite{Zic:2023gta,*Reardon:2023gzh}, CPTA DR1~\cite{Xu:2023wog}, and EPTA+InPTA DR2~\cite{Antoniadis:2023lym,*Antoniadis:2023ott} have revealed a common-spectrum signal that could potentially originate from GWs. Although the evidence for Hellings-Downs (HD) correlations~\cite{Hellings:1983fr}, a key signature of a GW-induced signal, varies among the different PTA data sets, the detection of a common-spectrum signal has generated significant interest in the scientific community~\cite{Bi:2023tib,*Chen:2023bms,*Chen:2023uiz,*Bi:2023ewq,*Wu:2023rib,*InternationalPulsarTimingArray:2023mzf}.
Various possible sources of the common-spectrum signal have been proposed, each with unique implications for fundamental physics and cosmology. Candidates include phase transitions in the early Universe~\cite{Witten:1984rs,*Hogan:1986qda,*Caprini:2010xv}, cosmic strings~\cite{Vilenkin:1981bx,*Damour:2001bk,*Damour:2004kw}, and domain walls~\cite{Vilenkin:1981zs,*Hiramatsu:2013qaa}, all of which are predicted to produce SGWBs with characteristic spectral shapes and amplitudes. Another intriguing possibility is the presence of scalar-induced gravitational waves (SIGWs)\cite{Ananda:2006af,*Baumann:2007zm,*Yuan:2019fwv,*Yuan:2019wwo,*Cai:2019bmk,*Chen:2024twp,*Chen:2024fir,*Liu:2023hpw,*Yi:2023mbm,*Fei:2023iel,*You:2023rmn,*Yi:2023npi}, which are generated by second-order perturbations of the scalar field during inflation \cite{Starobinsky:1979ty,*Mukhanov:1981xt,*Guth:1982ec,*Hawking:1982cz,*Starobinsky:1982ee,*Bardeen:1983qw}.

Inflation is a hypothetical period of exponential expansion in the early Universe that solves several cosmological problems, such as the horizon and flatness problems~\cite{Guth:1980zm,*Linde:1981mu,*Albrecht:1982wi,*Linde:1983gd}. During inflation, quantum fluctuations in the scalar field give rise to density perturbations, which later form the seeds for structure formation in the Universe~\cite{Mukhanov:1981xt,*Hawking:1982cz,*Starobinsky:1982ee,*Guth:1982ec,*Bardeen:1983qw}. These density perturbations also induce second-order perturbations in the metric, manifesting as SIGWs~\cite{Ananda:2006af,*Baumann:2007zm,*Saito:2008jc}. Interestingly, SIGWs can also accompany the formation of primordial black holes (PBHs)~\cite{Saito:2009jt,*Bugaev:2009zh,*Bugaev:2010bb,*Liu:2018ess,*Liu:2019rnx,*Huang:2024wse,*Chen:2024dxh,*Chen:2024joj,*Chen:2022qvg,*Liu:2020cds,*Liu:2020vsy,*Liu:2020bag,*Liu:2022cuj,*Chen:2018czv,*Yuan:2019udt,*Liu:2021jnw,*Yi:2023tdk,*Wu:2020drm,*Liu:2022wtq,*Meng:2022low}, which are hypothetical black holes that may have formed in the early Universe from the collapse of large density perturbations generated during inflation.
If the primordial curvature power spectrum exhibits a significant enhancement at certain scales, it can lead to the formation of PBHs, accompanied by a corresponding enhancement in the SIGW spectrum at the same scales. The properties of the inflationary model, such as the potential of the inflaton field and the duration of inflation, determine the characteristics of the primordial curvature power spectrum, which in turn influences the amplitude and spectral shape of the SIGWs~\cite{Maggiore:1999vm,*Boyle:2005se,*Watanabe:2006qe,*Bugaev:2009zh,*Bugaev:2010bb,*Kuroyanagi:2011fy,*Kuroyanagi:2014qza}. Studying SIGWs associated with PBH formation can provide valuable insights into the inflationary dynamics and the properties of the early Universe, as well as the potential existence of PBHs as a component of dark matter~\cite{Bird:2016dcv,*Sasaki:2016jop}.

In this work, we focus on SIGWs as a potential explanation for the common-spectrum signal detected by PTAs. We aim to constrain the inflationary parameters by comparing the predicted SIGW spectrum with the PTA data, assuming an equation of state $w=1/6$ and a sound speed $c_s=1$ following the end of inflation~\cite{Mukhanov:1990me,*Weinberg:2008zzc}. The $w=1/6$ case, arising from the $\phi^{14/5}$ potential, serves as a valuable example of how non-canonical kinetic terms \cite{Armendariz-Picon:2000ulo}, higher-order corrections \cite{Liddle:2003as}, effective field theory approaches \cite{Cheung:2007st}, and string theory/supergravity models \cite{Kachru:2003sx,*Baumann:2014nda} can lead to novel reheating dynamics and expand our understanding of the early Universe.
By considering this equation of state in our analysis and comparing the predicted SIGW spectrum with the PTA data, we can constrain the properties of the early Universe and investigate the potential origin of the common-spectrum signal detected by PTAs \cite{Domenech:2019quo,*Inomata:2019ivs,*Inomata:2019zqy,*Domenech:2020kqm,*Inomata:2020cck,*Fumagalli:2020nvq}. We calculate the corresponding SIGW spectrum using the formalism developed in~\cite{Ananda:2006af,*Baumann:2007zm} and compare it with the PTA data. Additionally, we investigate the impact of the reheating temperature, which marks the end of inflation and the beginning of the radiation-dominated era, on the SIGW spectrum~\cite{Kofman:1994rk,*Kofman:1997yn,*Bassett:2005xm,*Allahverdi:2010xz}.
This study provides valuable insights into the potential of PTAs to constrain inflationary models and shed light on the early Universe through the detection of SIGWs. The rest of this paper is organized as follows. Section \ref{sec:GW} describes the theoretical framework for SIGWs and the inflationary models considered in this work. Section \ref{sec:result} presents the methods used to analyze the PTA data and the results of our analysis. Finally, Section \ref{sec:con} discusses the implications of our findings.

\section{SIGWs and PBHs} 
\label{sec:GW}

SIGWs and PBHs are two fascinating phenomena that can arise from primordial scalar perturbations in the early Universe. These perturbations, thought to have originated from quantum fluctuations during cosmic inflation, can leave imprints on the Universe's evolution and structure. In this section, we explore the basic formulas and concepts related to SIGWs and PBHs, focusing on how primordial scalar perturbations can give rise to these phenomena and how their characteristics are influenced by various factors, such as the primordial power spectrum and the post-inflationary composition of the Universe. Understanding the theoretical foundations of SIGWs and PBHs is crucial for interpreting observational data and gaining insights into the early Universe and the nature of primordial fluctuations.

\subsection{Basic formulas for SIGWs}
SIGWs arise from the amplification of primordial scalar perturbations, which can potentially generate a detectable SGWB. The characteristics of the SIGW spectrum, including its amplitude and shape, are governed by several factors such as the power spectrum of primordial fluctuations and the post-inflationary composition of the Universe.
To model the primordial spectrum responsible for SIGWs, we employ a lognormal peak description that captures the enhancement of fluctuations on small scales. The primordial curvature power spectrum is parameterized as follows:
\begin{equation}
\label{PR}
\mathcal{P}_{\mathcal{R}} (k) = \frac{A}{\sqrt{2\pi}\Delta} \exp\! \left( -\frac{\ln^{2}(k/k_{*})}{2\Delta^{2}} \right),
\end{equation}
where $A$ represents the amplitude, $k_*$ denotes the characteristic scale, and $\Delta$ corresponds to the width of the spectrum. Our analysis focuses on the case of a narrow peak spectrum, where $\Delta \lesssim 0.1$, which is supported by our findings. In the limit of $\Delta \rightarrow 0$, the primordial power spectrum reduces to a $\delta$-function form, $\mathcal{P}_{\mathcal{R}} = A\, k\, \delta(k-k_*)$. The impact of a finite width on the SIGW spectrum can be expressed as~\cite{Pi:2020otn}:
\begin{equation}\label{eq:OmegaDelta}
\Omega_{\rm GW}h^2={\rm Erf}\!\left[\frac{1}{\Delta}\sinh^{-1}\frac{k}{2k_{\rm *}}\right]\Omega^\delta_{\rm GW}h^2,
\end{equation}
where $\Omega_{\rm GW}h^2$ represents the present-day GW energy density fraction with finite width, ${\rm Erf}$ denotes the error function, and $\Omega^\delta_{\rm GW}h^2$ corresponds to the GW spectrum induced by a $\delta$-function peak.

We adopt an instantaneous reheating scenario, assuming that the Universe transitions to the radiation-dominated era immediately when the mode with comoving wave number $k_\mathrm{rh}$ reenters the horizon. For Gaussian fluctuations and a constant equation of state parameter $w$ and sound speed $c_s$, a general integral formula for the SIGW spectrum can be derived. The present-day fraction of GW energy density is given by:
\begin{equation}\label{ogw0}
\begin{split}
\Omega_{\mathrm{GW}} h^2 \approx & 1.62 \times 10^{-5}\left(\frac{\Omega_{r, 0} h^2}{4.18 \times 10^{-5}}\right)   \left(\frac{g_{*r}(T_{\mathrm{rh}})}{106.75}\right)\left(\frac{g_{* s}(T_{\mathrm{rh}})}{106.75}\right)^{-\frac{4}{3}} \Omega_{\mathrm{GW}, \mathrm{rh}},    
\end{split}
\end{equation}
where $\Omega_{r,0}h^2$ denotes the present-day radiation fraction, $g_{*r}$ and $g_{*s}$ are the effective energy and entropy degrees of freedom~\cite{Saikawa:2018rcs}, respectively, and the subscript ``rh" refers to the time of instantaneous reheating.

The SIGW spectrum for scales $k \gtrsim k_{\rm rh}$ can be expressed as~\cite{Domenech:2021ztg}:
\begin{equation}
\Omega_{\mathrm{GW},\mathrm{rh}}=\left(\frac{k}{k_{\mathrm{rh}}}\right)^{-2 b} \int_0^{\infty} d v \int_{|1-v|}^{1+v} d u\, \mathcal{T}(u,v,b,c_s)\, \mathcal{P}_{\mathcal{R}}(k u)\, \mathcal{P}_{\mathcal{R}}(k v),
\end{equation}
where $b\equiv (1-3w)/(1+3w)$ and the transfer function $\mathcal{T}(u,v,b,c_s)$ is defined in terms of Ferrer's functions, Olver's function, and the Heaviside theta function. The transfer function can be further expressed using hypergeometric functions, as 
\begin{align}\label{eq:kernelaveragefinal}
{\cal T}(u,v,b,c_s)=& {\cal N}(b,c_s)\left(\frac{4v^2-(1-u^2+v^2)^2}{4u^2v^2}\right)^2{|1-y^2|^{b}} \Bigg\{\left(\mathsf{P}^{-b}_{b}(y)+\frac{b+2}{b+1}\mathsf{P}^{-b}_{b+2}(y)\right)^2\Theta(c_s(u+v)-1)\nonumber\\&
+\frac{4}{\pi^2}\left(\mathsf{Q}^{-b}_{b}(y)+\frac{b+2}{b+1}\mathsf{Q}^{-b}_{b+2}(y)\right)^2\Theta(c_s(u+v)-1)\nonumber\\&
+\frac{4}{\pi^2}\left({\cal Q}^{-b}_{b}(-y)+2\frac{b+2}{b+1}\mathsf{\cal Q}^{-b}_{b+2}(-y)\right)^2\Theta(1-c_s(u+v))\Bigg\},
\end{align}
where $\Theta$ is the Heaviside theta function, ${\cal Q}^{\mu}_{\nu}(x)$ is Olver's function, $y\equiv 1-[1-c_s^2(u-v)^2]/(2c_s^2 uv)$, $\mathsf{P}^{\mu}_{\nu}(x)$ and $\mathsf{Q}^{\mu}_{\nu}(x)$ are Ferrer's functions. The numerical coefficient is
\begin{align}
{\cal N}(b,c_s)\equiv \frac{4^{2b}}{3c_s^4}\Gamma(b+\tfrac{3}{2})^4\left(\frac{b+2}{2b+3}\right)^2\left(1+b\right)^{-2(1+b)}\,.
\end{align}

It is important to note that the SIGW spectrum, $\Omega_{\mathrm{GW},\mathrm{rh}}$, is proportional to $(k/k_{\rm rh})^2$ \cite{Domenech:2020kqm,Domenech:2021ztg,Liu:2023pau}, implying that the amplitude of the SIGW spectrum is suppressed by a factor of $(k_{\rm rh}/k_*)^2$. 
To establish the relation between the frequency $f$ and the temperature $T$, we can estimate that each mode with a wavenumber $k$ crosses the horizon at temperature $T$ using the approximation
\begin{equation}
\label{k-T}
    k\! \simeq \frac{\!1.5\!\times\!  10^7}{\mathrm{Mpc}}
    \left({g_{*r}(T)} \over {106.75} \right)^\frac{1}{2}\!\!
    \left({g_{*s}(T)} \over 106.75\right)^{-\frac{1}{3}}\!\!
    \left (\frac{T}{\rm GeV}\right),
\end{equation}
with the corresponding GW frequency $f$ at the current epoch for each mode $k$ given by
\begin{equation}
\label{k-f}
f = \frac{k}{2 \pi} \simeq 1.6\, \mathrm{nHz}
\left(\frac{k}{10^6\,\mathrm{Mpc}^{-1}}\right).
\end{equation}
Combining \Eq{k-T} and \Eq{k-f}, we obtain the relation between the frequency $f$ and the temperature $T$:
\begin{equation}
f \simeq 24\, \mathrm{nHz} \left(\frac{g_{*r}(T)}{106.75}\right)^{\frac{1}{2}}
\left(\frac{g_{*s}(T)}{106.75}\right)^{-\frac{1}{3}}
\left(\frac{T}{\mathrm{GeV}}\right).
\end{equation}
The lower bound on the reheating temperature for Big Bang Nucleosynthesis (BBN) is $T_{\mathrm{rh}} \geq 4\,\mathrm{MeV}$ \cite{Kawasaki:1999na,*Kawasaki:2000en,*Hannestad:2004px,*Hasegawa:2019jsa}, which translates to constraints on the reheating frequency $f_{\mathrm{rh}}$ for the reheating mode $k_{\mathrm{rh}}$ as $f_{\mathrm{rh}} \gtrsim 0.1\,\mathrm{nHz}$. This frequency range lies below the sensitivity of current PTAs.

\subsection{Basic formulas for PBHs}

PBHs are formed through gravitational collapse when the density contrast $\delta\rho/\rho$ surpasses a threshold $\delta_c$ in Hubble patches, depending on the equation of state parameter $w$ at the moment of re-entry~\cite{Zeldovich:1967lct,*Hawking:1971ei,*Carr:1974nx,*Meszaros:1974tb,*Carr:1975qj,*Musco:2004ak,*Musco:2008hv,*Musco:2012au,*Harada:2013epa,*Escriva:2020tak}. To ensure a conservative estimation, we adopt Carr's criterion in the uniform Hubble slice~\cite{Carr:1975qj}, which can be translated to the comoving slice as follows~\cite{Domenech:2020ers,*Balaji:2023ehk}:
\begin{align}\label{dcb}
\delta_c \simeq \frac{3(1+w)}{5+3w} c_s^2=\frac{3(1+w)}{5+3w}=\frac{7}{11}.
\end{align}
The choice of the critical density threshold $\delta_c$ is justified by the fact that the fluctuations of the scalar field propagate with a speed of sound squared. This property makes the formation of PBHs more challenging compared to an adiabatic perfect fluid. Consequently, we adopt the upper limit of the density threshold as suggested by Ref.~\cite{Harada:2013epa}. It is worth noting that the case of general $c_s^2=w$ has been investigated in Ref.~\cite{Musco:2012au}.

The formation of PBHs is a complex process that depends on the interplay between the density fluctuations and the background cosmological evolution. The density contrast $\delta\rho/\rho$ quantifies the magnitude of the density fluctuations relative to the average density of the Universe. When this contrast exceeds the critical threshold $\delta_c$, the gravitational force overcomes the pressure support within a Hubble patch, leading to the collapse of the overdense region and the formation of a PBH. The abundance of PBHs at the time of their formation can be determined using the Press-Schechter theory~\cite{Sasaki:2018dmp} as
\begin{equation}
    \beta(M)=\frac{\gamma}{2}\text{erfc}\left(\frac{\delta_c(w)}{\sqrt2\sigma(M)}\right),
\end{equation}
where $\gamma\approx0.2$ represents the fraction of matter within the Hubble horizon that undergoes collapse to form PBHs. The quantity $\sigma(M)$ denotes the variance of the density perturbation smoothed over the mass scale of $M$. Using the relation between curvature perturbation $\mathcal{R}$ and density contrast $\delta$,
\begin{equation}
    \delta(t, q)=\frac{2(1+w)}{5+3 w}\left(\frac{q}{a H}\right)^2 \mathcal{R}(q)=\frac{14}{33} \left(\frac{q}{a H}\right)^2 \mathcal{R}(q), 
\end{equation}
we can compute $\sigma(M)$ as
\begin{equation}
\label{sigma}
\sigma^{2} =\left(\frac{14}{33}\right)^{\!\!2}\!\!\! \int_{0}^{\infty}\! \frac{\mathrm{d}q}{q}\, \tilde{W}^{2}(\frac{q}{k})\left(\frac{q}{k}\right)^{4} T^{2}(\frac{q}{k}) \mathcal{P}_{\mathcal{R}}(q),
\end{equation}
where $\tilde{W}(q/k)=\exp(-q^{2}/k^{2}/2)$ is the Gaussian window function and $T(q/k)$ is the transfer function that describes the evolution of density perturbations from the time of horizon entry to the time of PBH formation. The masses of PBHs are connected to the comoving scale by
\begin{align}\label{mpbh}
\frac{M_{\rm PBH}}{M_\odot}&\approx  0.01\frac{\gamma}{0.2}\!\left(\frac{k_{\rm rh}}{k_*}\right)^{\!\!\!\frac{3(1+w)}{1+3w}}\!\!\! \left(\frac{106.75}{g_{*r}(T_{\rm rh})}\right)^{\!\!\frac12}\!\!\!\left(\frac{{\rm GeV}}{T_{\rm rh}}\right)^{\!\!\! 2}\approx 0.01\frac{\gamma}{0.2}\!\left(\frac{k_{\rm rh}}{k_*}\right)^{\!\!\!\frac{7}{3}}\!\!\! \left(\frac{106.75}{g_{*r}(T_{\rm rh})}\right)^{\!\!\frac12}\!\!\!\left(\frac{{\rm GeV}}{T_{\rm rh}}\right)^{\!\!\! 2}.
\end{align}
In the case of a very sharp peak in the primordial scalar power spectrum, it results in a monochromatic mass function for the PBHs, meaning that the PBHs formed have a narrow range of masses. The abundance of PBHs, expressed as the PBH energy fraction with respect to cold dark matter, is described by
\begin{equation}\label{fpbh}
f_{\rm PBH}\equiv\frac{\Omega_\text{PBH}}{\Omega_\text{CDM}} \approx 1.5\times 10^{13}\beta\left(\frac{k_*}{k_{\rm rh}}\right)^{\frac{2}{3}}\left(\frac{T_{\rm rh}}{{\rm GeV}}\right) \left(\frac{g_{*s}(T_{\rm rh})}{106.75}\right)^{-1}\left(\frac{g_{*r}(T_{\rm rh})}{106.75}\right),
\end{equation}
where $\Omega_\text{PBH}$ and $\Omega_\text{CDM}$ are the current energy densities of PBHs and cold dark matter, respectively.

\section{Methodology and Results}\label{sec:result}

In this section, we explore the hypothesis that the signal detected in the NANOGrav 15-year data set originates from SIGWs generated by a primordial curvature power spectrum with a finite-width lognormal form, as described in Equation \eqref{PR}. To test this hypothesis and constrain the parameters of the primordial curvature power spectrum and the reheating temperature $T_{\rm rh}$, we employ Bayesian inference techniques on the NANOGrav 15-year data set.

\begin{table}[tbp]
\centering
\begin{tabular}{l|llllll}
\hline
\hline
Parameter & $\log_{10} (f_*/\mathrm{Hz})$ \quad & $\log_{10} \Delta$ \quad & $\log_{10} A$ \quad & $\log_{10}(T_{\rm rh}/{\rm GeV})$\\
\hline
prior& $\mathcal{U}(-10, -2)$ & $\mathcal{U}(-6, -1)$ & $\mathcal{U}(-5, 1)$ & $\mathcal{U}(-2.5, 0.6)$\\
\hline
posterior & $-6.14^{+1.23}_{-1.21}$ & $-3.55^{+2.20}_{-2.15}$& $-0.27^{+1.17}_{-1.21}$ & $-0.35^{+0.80}_{-1.72}$\\
\hline 
\hline 
\end{tabular}
\caption{
Summarizes on the prior distributions, posterior medians, and 90\% credible intervals for the reheating temperature ($T_{\rm rh}$) and the parameters characterizing the primordial curvature power spectrum (Equation \eqref{PR}). These results were obtained by performing a Bayesian analysis on the NANOGrav 15-year data set, employing uniform prior distributions ($\mathcal{U}$) for all parameters. The posterior medians indicate the most likely values given the data and the model, while the 90\% credible intervals represent the range of plausible values consistent with the observations. The analysis reveals a strong preference for a narrow peak in the primordial power spectrum ($\Delta < 0.001$) and a lower limit on the reheating temperature ($T_{\rm rh} \geq 0.1 {\rm GeV}$) that agrees with constraints from BBN.
}
\label{Tab:prior}
\end{table}


Our analysis begins by calculating the energy density of SIGWs at the frequencies corresponding to the 14 bins provided in the NANOGrav 15-year data set \cite{NANOGrav:2023gor,*NANOGrav:2023hvm}. We then construct logarithmic probability density functions for each bin using kernel density estimation. The overall likelihood function is obtained by multiplying the probabilities from all 14 bins \cite{Moore:2021ibq,*Lamb:2023jls,*Liu:2023ymk,*Wu:2023hsa,*Jin:2023wri,Liu:2023pau}:
\begin{equation}
\mathcal{L}(\Lambda)=\prod_{i=1}^{14} \mathcal{L}_i\left(\Omega_{\mathrm{GW}}\left(f_i, \Lambda \right)\right),
\end{equation}
where $\Lambda = \{A, \Delta, f_*, T_{\rm rh}\}$ represents the set of model parameters, including those related to the primordial curvature power spectrum (Equation \eqref{PR}) and the reheating temperature $T_{\rm rh}$. We assign uninformative uniform priors to these parameters, as listed in Table \ref{Tab:prior}.
To compute the posterior distribution of the model parameters, we utilize the \texttt{Bilby} code \cite{Ashton:2018jfp}, which implements the \texttt{dynesty} nested sampling algorithm. This allows us to efficiently explore the high-dimensional parameter space and obtain reliable estimates of the posterior distributions.

\begin{figure}[thbp!]
	\centering
 \includegraphics[width=0.8\textwidth]{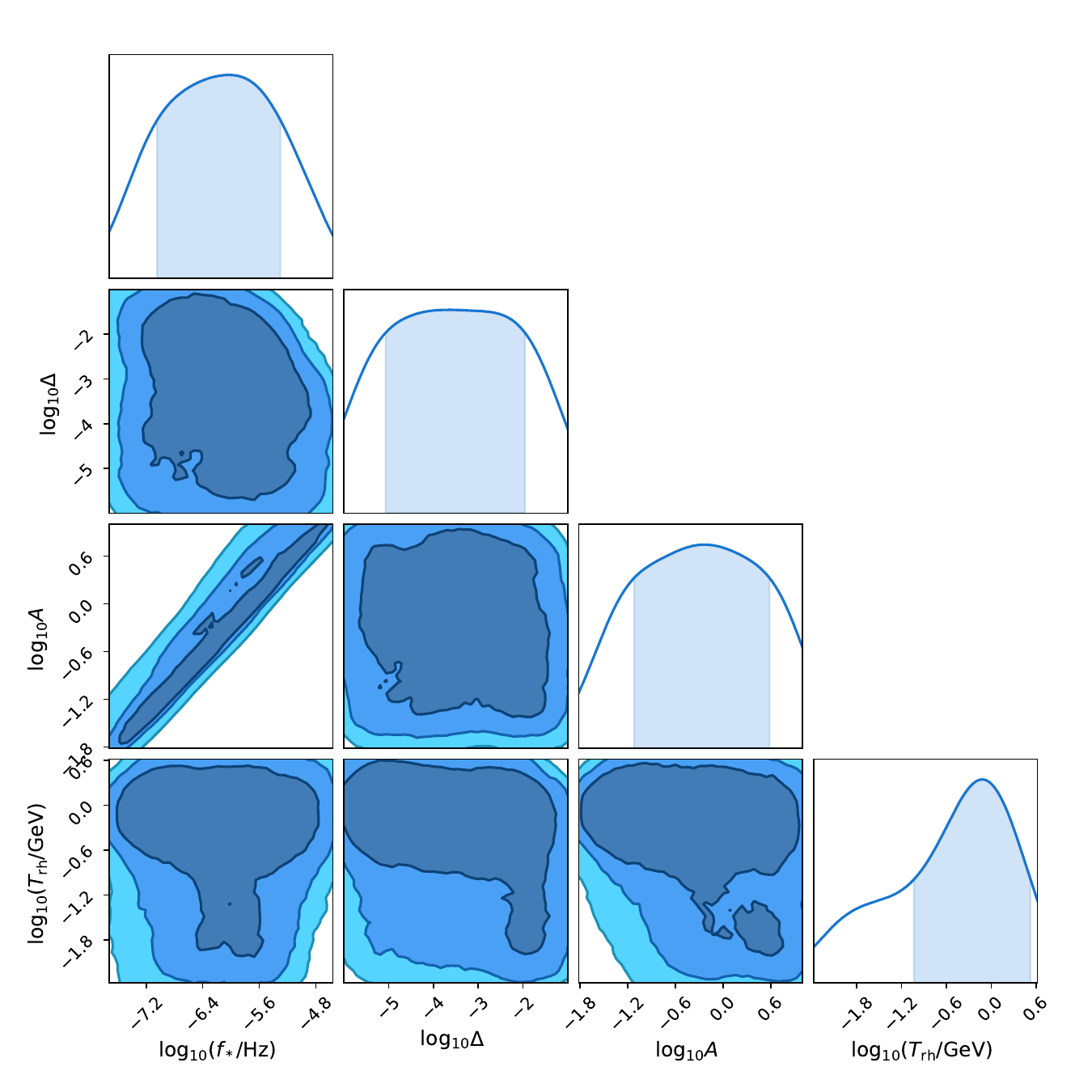}
	 \caption{\label{fig:post}
Results of our Bayesian inference analysis, which aims to constrain the parameters of the primordial curvature power spectrum (Eq.\eqref{PR}) and the reheating temperature ($T_{\rm rh}$) using the NANOGrav 15-year data set. The diagonal panels showcase the marginalized one-dimensional posterior probability distributions for each parameter, providing insights into their individual constraints. The off-diagonal panels display the two-dimensional joint posterior distributions, revealing the correlations and dependencies between pairs of parameters. The relationship between the peak scale $k_*$ in the primordial power spectrum and the corresponding frequency $f_*$ in the posterior distributions is determined by Equation \eqref{k-f}, which accounts for the evolution of primordial perturbations and their impact on the observed gravitational wave spectrum. These posterior distributions encapsulate the probabilistic constraints on the parameters, given the assumed model and the observational data. They allow us to identify the most probable parameter values, quantify the uncertainties associated with these estimates, and draw conclusions about the shape and amplitude of the primordial curvature power spectrum, as well as the timing and conditions of the reheating phase in the early Universe.
}
\end{figure}

\begin{figure}[thbp!]
	\centering
 \includegraphics[width=0.8\textwidth]{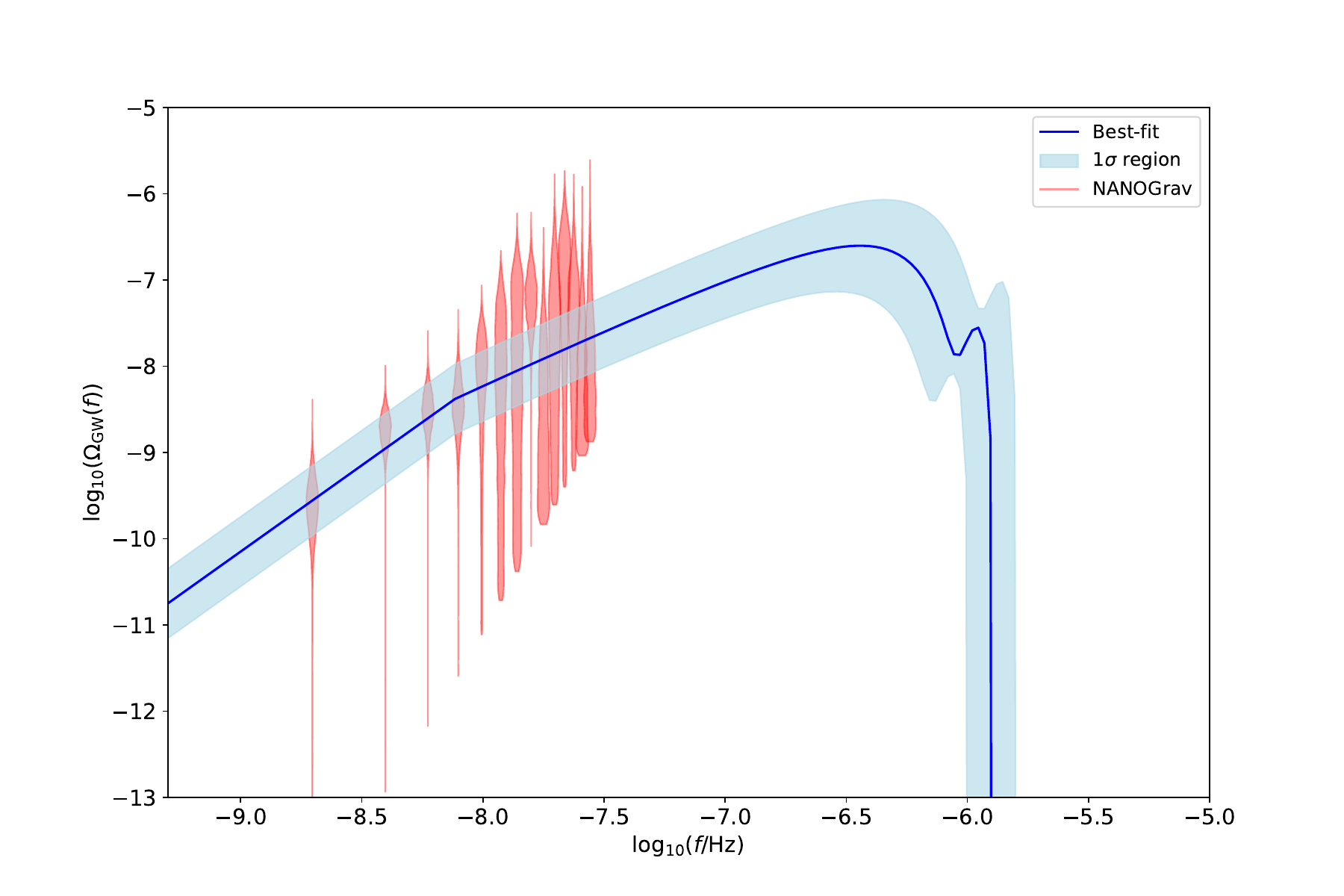}
	\caption{\label{fig:ogw}
 The energy density spectrum of SIGWs calculated using the best-fit parameter values obtained from our analysis. The blue curve represents the predicted SIGW energy density spectrum, while the blue region denotes the 1-$\sigma$ confidence interval around this spectrum. For comparison, the red violin plots show the energy density estimates of the free spectrum derived from the NANOGrav 15-year data set, with the width of each violin indicating the probability density of the energy density at the corresponding frequency. The presence of a distinct turning point in the SIGW spectrum around $f \sim 10^{-8.1} ~ {\rm Hz}$ is attributed to a change in the equation of state parameter $w$, which significantly influences the evolution of SIGWs. Additionally, the sharp decrease in the SIGW energy density at high frequencies ($f \sim 10^{-6} ~ {\rm Hz}$) is a direct consequence of the narrow peak in the primordial curvature power spectrum, as required by the NANOGrav data.
By comparing the predicted SIGW spectra with the observational constraints, this figure demonstrates the agreement between our best-fit models and the available data, supporting the hypothesis that the NANOGrav signal originates from SIGWs generated by a primordial curvature power spectrum with a narrow peak.
 }
\end{figure}

\begin{figure}[thbp!]
	\centering
 \includegraphics[width=0.8\textwidth]{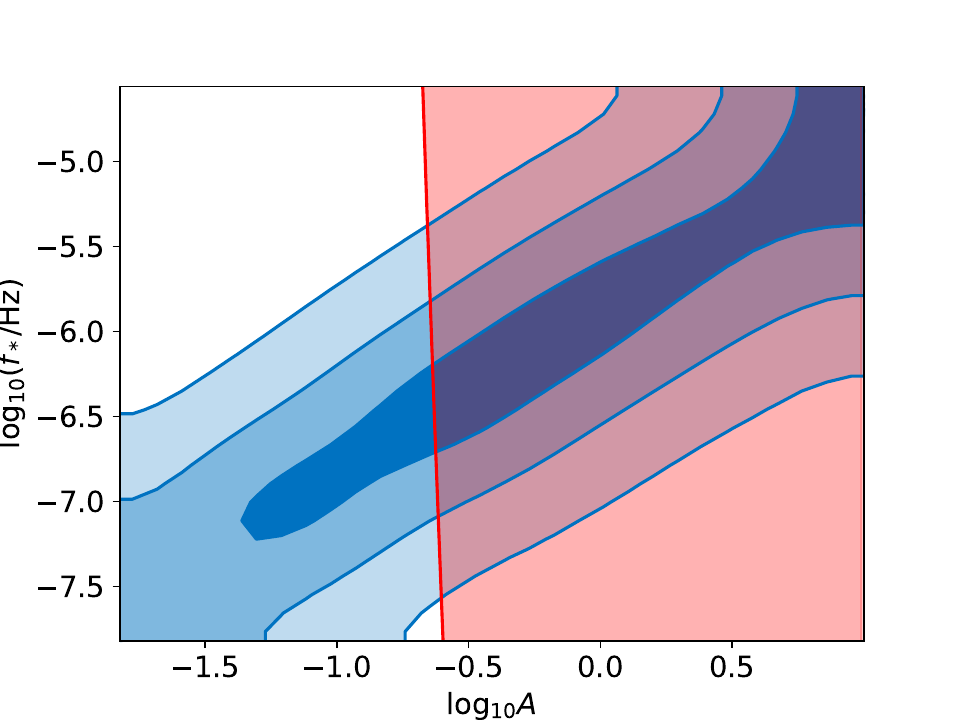}
	\caption{\label{fig:post_logA_logfstar}
 Two-dimensional posterior distribution for the amplitude ($A$) and peak frequency ($f_*$) parameters of the primordial curvature power spectrum, with the blue contours representing the $1\sigma$, $2\sigma$, and $3\sigma$ credible regions. The red shaded area indicates the region of parameter space excluded due to the overproduction of PBHs, where the PBH abundance exceeds unity ($f_{\rm PBH} > 1$). This analysis assumes fixed values for the other model parameters, namely the width of the power spectrum peak ($\log_{10} \Delta=-3.55$) and the reheating temperature ($\log_{10} (T_{\mathrm{rh}}/\mathrm{GeV}) = -0.35$). The posterior distribution and the excluded region provide valuable insights into the constraints on the primordial power spectrum parameters and the formation of PBHs, highlighting the importance of considering PBH abundance limits when studying the early Universe and the generation of SIGWs.
 }
\end{figure}

Figure \ref{fig:post} presents the resulting posterior distributions for the parameters of the primordial curvature power spectrum and the reheating temperature. The median posterior values and the 90\% credible intervals for these parameters are summarized in Table \ref{Tab:prior}. It is important to note that the scale $k_*$ and the corresponding frequency $f_*$ in the posterior distributions are related by Equation \eqref{k-f}.
The NANOGrav 15-year data set imposes tight constraints on the width of the primordial curvature power spectrum, requiring $\Delta \leq 0.001$ at a 90\% confidence level. This result supports the narrow peak assumption made in Section \ref{sec:GW} and has important implications for the shape of the primordial curvature power spectrum. Moreover, the lower bound on the reheating temperature, $T_{\rm rh} \geq 0.1 {\rm GeV}$, is consistent with the constraints from BBN, ensuring that our model is compatible with established cosmological observations. The constraints on the amplitude $A$ and the peak frequencies $f_*$ of the primordial curvature power spectrum agree with those reported by the NANOGrav Collaboration in their study of new physical phenomena \cite{NANOGrav:2023hvm}.

Using the optimal parameter values obtained from the posterior distributions (Table \ref{Tab:prior}), we calculate the energy density spectrum of SIGWs. Figure \ref{fig:ogw} shows the resulting spectrum (blue line) along with the 1-$\sigma$ confidence intervals (blue region). Interestingly, we observe a distinct turning point in the energy density spectrum around a frequency of $f \sim 10^{-8.1} ~ {\rm Hz}$. This feature corresponds to a change in the equation of state parameter $w$, which significantly influences the evolution of SIGWs. The presence of this turning point in the SIGW spectrum provides a unique signature that can be used to probe the properties of the early Universe and the nature of the primordial curvature perturbations.

Another notable feature in the SIGW energy density spectrum is the sharp decrease in the ultraviolet region around $f \sim 10^{-6} ~ {\rm Hz}$. This characteristic is a direct consequence of the narrow peak in the primordial curvature power spectrum, as required by the NANOGrav data ($\Delta < 0.001$). In the limiting case of a $\delta$-function form of the power spectrum, $\mathcal{P}_\mathcal{R}(k)=A\delta[\ln(k/k_*)]$, the energy density of SIGWs is expected to cut off at scales $k \geq 2 k_*$ \cite{Kohri:2018awv}. Our analysis demonstrates that the NANOGrav data favors a primordial curvature power spectrum with a very narrow peak, leading to the observed sharp decrease in the SIGW energy density spectrum.

To further investigate the implications of our model, we explore the parameter space of the amplitude $A$ and the peak frequency $f_*$. Figure \ref{fig:post_logA_logfstar} shows the two-dimensional posterior distribution for these parameters, with the red region indicating the excluded parameter space due to the overproduction of PBHs, i.e., $f_{\rm PBH} > 1$. This result confirms the previously known PBH overproduction issue \cite{Balaji:2023ehk} and emphasizes the importance of considering PBH constraints when studying SIGWs and the primordial curvature power spectrum.

The excluded region in the $A$-$f_*$ parameter space provides valuable insights into the formation of PBHs and their relation to the primordial curvature perturbations. The overproduction of PBHs occurs when the amplitude of the primordial curvature power spectrum is sufficiently large, leading to the collapse of overdense regions into black holes. By simultaneously accounting for the observational constraints from the NANOGrav data and the theoretical limits on PBH abundance, our analysis enables a more comprehensive understanding of the viable parameter space for the primordial curvature power spectrum and its implications for the formation of PBHs and the generation of SIGWs.

\section{Conclusion and discussion}
\label{sec:con}

The stochastic signal observed by collaborations such as NANOGrav, PPTA, EPTA+InPTA, and CPTA likely originates from GWs induced by primordial curvature perturbations generated during inflation. Following the inflationary epoch, the Universe undergoes a reheating phase, which aims to increase its temperature and restore the hot Big Bang scenario. Crucial information about the small-scale properties of inflation and the subsequent reheating process can be encoded in the energy density of the SGWB. This study explores the small-scale characteristics of inflation and reheating, focusing on the specific case where the equation of state during the reheating phase is characterized by $w=1/6$.

To model the power spectrum of the primordial curvature perturbations, we adopt a lognormal form. This functional form is chosen for its flexibility and ability to capture various features in the power spectrum. By applying Bayesian methods and analyzing the NANOGrav 15-year data set, we obtain posterior distributions for the key parameters that characterize the lognormal power spectrum and the reheating process. The obtained posterior values for these parameters are $\log_{10} (f_*/\mathrm{Hz}) = -6.14^{+1.23}_{-1.21}$,  $\log_{10} \Delta = -3.55^{+2.20}_{-2.15}$,  $\log_{10} A= -0.27^{+1.17}_{-1.21}$,  $\log_{10}(T_{\rm rh}/{\rm GeV}) = -0.35^{+0.80}_{-1.72}$. These results provide valuable insights into the shape of the primordial curvature power spectrum and the energy scale of reheating.

In summary, the lognormal modeling of the primordial curvature power spectrum, combined with the analysis of the NANOGrav 15-year data set, has yielded important constraints on the key parameters describing the shape of the power spectrum and the energy scale of reheating. The distinctive turning point in the SGWB spectrum, corresponding to the transition from reheating to the radiation-dominated era, offers a promising avenue for further investigation. As future observational data on SIGWs become available, we can expect to gain even deeper insights into the fundamental aspects of cosmological evolution during the early Universe, potentially revolutionizing our understanding of inflation, reheating, and the origin of the Universe as we know it.

\section*{Acknowledgments}
ZQY is supported by the National Natural Science Foundation of China under Grant No. 12305059. FGZ is supported by the National Natural Science Foundation of China under Grant No. 12305075.

\bibliography{ref}
\end{document}